\documentclass[5p]{elsarticle}
\usepackage[utf8x]{inputenc}
\usepackage{color}
\usepackage{amsmath,amscd,amsfonts,amssymb}
\usepackage{hyperref}
\definecolor{darkblue}{rgb}{0.0,0.0,0.3}
\hypersetup{colorlinks,breaklinks,
linkcolor=darkblue,urlcolor=darkblue,
anchorcolor=darkblue,citecolor=
darkblue,pdfauthor=Arvind, pdftitle=Weak-Discord}
\DeclareMathOperator{\sech}{sech}

\newcommand{\ie}{{\it i.~$\!$e.~}}
\journal{Physics Letters A}
\begin{document}
\begin{frontmatter}
\title{Simulating the effect of weak measurements by a phase damping 
channel and determining different measures of 
bipartite correlations in nuclear magnetic
resonance}
\author[iiserm]{Akanksha Gautam}
\ead{akankshagautam@iisermohali.ac.in}
\author[um]{Varad R.~Pande}
\ead{varadrpande@gmail.com}
\author[iiserm,china]{Amandeep Singh}
\ead{singh@sustech.edu.cn}
\author[iiserm]{Kavita Dorai}
\ead{kavita@iisermohali.ac.in}
\author[iiserm]{Arvind\corref{cor}\fnref{fn}}
\ead{arvind@iisermohali.ac.in}
\cortext[cor]{Corresponding Author}
\fntext[fn]{Tel \& Fax:+91-172-2240266}
\address[iiserm]{Department of Physical Sciences, Indian
Institute of Science Education \& 
Research Mohali, Sector 81 SAS Nagar, 
Manauli PO 140306 Punjab India.}
\address[um]{Department of Physics, University of Maryland,
Baltimore County (UMBC), Baltimore, Maryland 21250, USA.}
\address[china]{Shenzhen Institute for Quantum Science and
Engineering, and Department of Physics,
Southern University of Science and Technology, Shenzhen 518055, China}
\begin{abstract}
Quantum discord is a measure based on local projective
measurements which captures quantum correlations that
may not be fully
captured by entanglement.  A change in the measurement process,
achieved by replacing rank-one projectors with a weak
positive operator-valued measure (POVM), allows one to
define  
weak variants of quantum discord.  
In this work, we experimentally simulate the effect of
a weak POVM on a nuclear magnetic resonance quantum
information processor. 
The two-qubit system under investigation is part of a three-qubit 
system, where one of the qubits is used as an ancillary 
to implement the phase damping channel.  The strength
of the weak POVM is controlled by varying the strength of
the phase damping channel. We experimentally observed 
two weak variants of quantum discord namely,
super quantum discord and weak quantum discord,
in two-qubit Werner and Bell-diagonal
states. 
The resultant dynamics of the states is
investigated as a function of the measurement strength.
\end{abstract} 
\end{frontmatter}
\section{Introduction} 
Quantum correlations play an important role in quantum
communication and quantum information processing
\cite{nielsen-book-02}. While quantum entanglement was
discovered early on by Schr{\"o}dinger \cite{schrod-pps-35},
Ollivier and Zurek \cite{ollivier-prl-01} and Henderson and
Vedral \cite{henderson-iop-2001}, independently pointed out
that a different type of quantum correlations can exist
in bipartite systems.
The measure to quantify such
correlations is termed as quantum discord
(QD)~\cite{ollivier-prl-01,henderson-iop-2001}. The
evaluation of QD is computationally a hard task as it
involves numerical optimization and hence, alternative
measures have been proposed such as geometric quantum
discord~\cite{daki-prl-10}, Gaussian geometric discord in
continuous variable systems~\cite{adesso-prl-10} and
relative entropy of discord~\cite{modi-prl-10}. The presence
of nonclassical correlations in bipartite states was
measured experimentally on a nuclear magnetic resonance
(NMR) setup~\cite{silva-pra-13}, witnessed via a single-shot
experiment~\cite{Asingh-pra-17} and its preservation has
also been explored using NMR~\cite{singh-epl-17}. Several
recent studies have explored the advantage of nonclassical
correlations in quantum information processing even if the
state has almost null
entanglement~\cite{pirandola-sr-14,daki-np-12,modi-rmp-12}.

In classical information theory, the mutual information
between two random variables has two equivalent expressions
which can be computed from the respective Shannon entropies.
In contrast, for quantum mutual information, these two
definitions give rise to different values as one of the
expressions requires the von Neumann entropy conditioned by
projective measurements. QD is defined as the difference
between total mutual information and classical mutual
information, wherein classical mutual information is found
by determining the possible
information gain about one subsystem while measuring the
other subsystem via projective
measurements~\cite{henderson-iop-2001}. What if one replaces
these projective measurements by weak measurements? The crux
of a weak measurement lies in the weak interaction between
the system to be measured and the measuring apparatus.
While originally the weak measurements were used in the
context of weak values~\cite{aharonov-prl-88}, it was later
realized that any projective measurement can be obtained by
a sequence of weak measurements~\cite{oreshkov-prl-05}.
These ideas have been used in several applications including
the observation of spin Hall effect of
light~\cite{hosten-science-08}, direct measurement of single
photon wavefunction~\cite{lundeen-nature-11}, protection of
quantum entanglement from decoherence~\cite{kim-np-11},
feedback control of a quantum system~\cite{gillett-prl-10},
to amplify small transverse deflections of an optical beam
in a Sagnac interferometer~\cite{dixon-prl-09} and for
quantum state
tomography~\cite{lundeen-prl-12,wu-sr-13,lundeen-prl-16}.

Weak measurements allow us to define two different
variants of quantum discord, which stem from the fact that
QD can be defined in two mathematically equivalent
ways~\cite{rulli}. 
Weak measurements have been
utilized in a new definition of bipartite correlations
through a weak variant of QD termed super quantum discord
(SQD)~\cite{singh-aop-14,pande-pla-2017}, which is
numerically greater than QD. Since then
SQD has attracted interest in various
contexts in quantum information
processing~\cite{wang-qip-14,li-ijtp-2015,jing-qip-17,
mirmasoudi-jpamt-18}.  Recently, the question of
whether weak measurements can be used to gain more
information about correlations present in a quantum system,
was explored by formulating another weak variant of QD
termed weak quantum discord (WQD), which never exceeds
QD~\cite{xiang,dieguez-qip-18}. While projective
measurements result in a loss of coherence, the precise
relationship between decoherence and measurement needs to be
clarified.  Simulations of the phase damping (PD) channel
have been realized as projective measurements in
NMR~\cite{lee-apl-06} and using this method quantum
teleportation and the quantum Zeno effect have been
implemented in an NMR setup~\cite{xiao-pla-06}.

In this work we experimentally 
implement the effect
of a weak POVM via a phase damping channel, in two-qubit
states.  The PD channel is a non-unitary operation used to
model the decoherence process~\cite{nielsen-book-02}.
We have also investigated the experimental behavior of
two weak variants of QD namely, SQD and WQD. The
determination of SQD and WQD from a
bipartite state requires a weak measurement on one subsystem
instead of a projective measurement.  The result of a weak
measurement on a Bell-diagonal state and on a Werner state
while finding SQD and WQD, affects the state in the same
manner as if a PD channel was acting on it.  We mapped the
weak measurement to a PD channel and the weak measurement
strength is controlled by tuning the strength of the PD
channel.  We use three nuclear spins to encode three qubits,
where the two qubits are used to prepare two-qubit states
and the third qubit acts as an ancillary qubit used to
simulate a PD channel acting on one of the qubit of
two-qubit states~\cite{xin-pra-17}.  This study reports the
first experimental implementation of  controlled weak
measurements on an NMR hardware.  We successfully
demonstrated that both SQD and WQD approach QD as the
measurement strength is increased.

This article is organized as
follows:~Sec.~\ref{theoretical-description-QD} describes
nonclassical correlations as quantified by QD.
Sec.~\ref{sqd-theory} and Sec.~\ref{wqd-theory}
describes weak measurement and its application in
determining the SQD and WQD respectively, in two-qubit
states. 
Sec.~\ref{phase_damping_and_weak_POVM}
details the mapping of weak measurement to the PD channel
and Sec.~\ref{Experiment} contains experimental results of
implementing a PD channel to observe SQD and WQD.
Sec.~\ref{conclusion} contains some concluding remarks.
\section{Quantum Discord and Weak Measurements}
\subsection{Quantum Discord} 
\label{theoretical-description-QD} It was independently
noted by Ollivier and Zurek~\cite{ollivier-prl-01} and by
Henderson and Vedral~\cite{henderson-iop-2001} that certain
types of mixed separable states have zero entanglement, yet
contain nonclassical correlations. 
Mathematically,
QD captures these correlations and is the difference between
the total correlation $I(\rho_{AB})$ and the classical correlation
$J(\rho_{A\vert B})$:
\begin{eqnarray}
\label{TC}
I(\rho_{AB})&=&S(\rho_{A})+S(\rho_{B})-S(\rho_{AB})
\nonumber \\
J(\rho_{A\vert B})&=&S(\rho_{A})-S(\rho_{A\vert B})
\label{CCo}
\end{eqnarray}
where $S(\rho_{AB})=-\mathrm{Tr}(\rho_{AB}\log_{2}
\rho_{AB})$  is the von Neumann entropy of the quantum
state,
$\rho_{AB}$ is shared between two parties $A$ and $B$,
$\rho_{A,B}=\mathrm{Tr}_{B,A}(\rho_{AB})$ is the reduced
density operator of a subsystem, and $S(\rho_{A\vert B})$ is
the conditional von Neumann entropy of subsystem $A$ when
subsystem $B$ has already been measured.
Quantum discord is then defined as: 
\begin{equation}
QD(\rho_{AB})= I(\rho_{AB})-\max J(\rho_{A\vert B})
\label{QD}
\end{equation}
where the maxima is taken over all possible projective
measurements on the subsystem $B$.

It turns out that we can also write quantum discord in a
mathematically equivalent form~\cite{rulli}:
\begin{equation}
QD(\rho_{AB})= I(\rho_{AB})-\max I(\rho^{\prime}_{AB})
\label{QD_alt}
\end{equation}
where $\rho^{\prime}_{AB}$ is the density operator of the
combined system after a projective measurement over
subsystem $B$ has
been carried out. As has been discussed in a lucid manner
in~\cite{dieguez-qip-18} this alternative mathematical form that was 
originally introduced for defining QD for  multipartite
systems, also leads to an alternative interpretation of QD and
is useful in generalizing QD to the weak measurement
scenario.

For a two-qubit system, projective measurements on the 
single-qubit subsystem $B$ can be
characterized by the Bloch sphere direction $\theta,\phi$.
The corresponding  
projectors $\Pi_{1}^{\theta,\phi}$ and
$\Pi_{2}^{\theta,\phi}$ can be constructed utilizing two
orthogonal vectors as follows:
\begin{eqnarray}
&\vert\psi\rangle_{1}^{\theta,\phi}=
\cos\frac{\theta}{2}\vert0\rangle+e^{i\phi}\sin\frac{\theta}{2}\vert
1\rangle 
\nonumber\\
&\vert\psi\rangle_{2}^{\theta,\phi}=
-\sin\frac{\theta}{2}\vert0\rangle+e^{i\phi}\cos\frac{\theta}{2}\vert
1\rangle 
\nonumber\\
&\Pi_{1}^{\theta,\phi}=
\vert\psi\rangle_{1}^{\theta,\phi} 
\langle\psi\vert_{1}^{\theta,\phi}
\quad {\rm and}\quad
\Pi_{2}^{\theta,\phi}=
\vert\psi\rangle_{2}^{\theta,\phi}
\langle\psi\vert_{2}^{\theta,\phi}.
\label{orth_proj}
\end{eqnarray}

The state of the subsystem $A$ after a projective measurement on
subsystem $B$ gives a positive result for the projector
$\Pi_j^{\theta,\phi}$ and can be written as:
\begin{equation}\label{state_after_meas}
\rho_{A\vert
\Pi_{j}^{\theta,\phi}}=\dfrac{1}{p_{j}}\mathrm{Tr}_{B}\left[\left(I\otimes
\Pi_{j}^{\theta,\phi}\right)\rho_{AB}\left(I\otimes
\Pi_{j}^{\theta,\phi}\right)\right]
\end{equation}
where $p_{j}$ is the probability of the measurement outcome
corresponding to the projectors $\Pi_{j}^{\theta,\phi}$.
The state of the combined system after a measurement on the
system $B$ can be written as:
\begin{equation}
\rho^{\prime}_{AB}= \sum_{j=1}^{2} \left[\left(I\otimes
\Pi_{j}^{\theta,\phi}\right)\rho_{AB}\left(I\otimes
\Pi_{j}^{\theta,\phi}\right)\right].
\end{equation}
The conditional von Neumann entropy required in Eq.~(\ref{CCo})
is given by:
\begin{equation}
\label{Cond_entropy} S(\rho_{A\vert
B})=\sum_{j=1}^{2}p_{j}S(\rho_{A\vert
\Pi_{j}^{\theta,\phi}}). 
\end{equation}
QD can be computed by utilizing the above defined projectors in
Eq.(\ref{Cond_entropy}), followed by substituting the
parameterized conditional entropy into
Eq.(\ref{CCo}): 
\begin{subequations}
\begin{eqnarray}
QD(\rho_{AB})
&=&I(\rho_{AB})-\max_{\lbrace
\theta,\phi \rbrace}J(\rho_{A\vert B}) 
\label{QD2_1}
\\
&=&I(\rho_{AB})-\max_{\lbrace
\theta,\phi \rbrace}I(\rho^{\prime}_{AB}). 
\label{QD2_2} 
\end{eqnarray}
\end{subequations}
Thus one can compute QD in two different ways: by maximizing
the $J(\rho_{A\vert B})$ or by maximizing
$I(\rho^{\prime}_{AB})$ (which for projective measurements
turn out to be the same) over $\theta\in [0,\pi]$ and
$\phi\in[0,2\pi]$, and substituting $\theta$ and $\phi$ back
into Eqs.~(\ref{QD2_1}) \&~(\ref{QD2_2}). 
\subsection{Weak Variants of Quantum Discord} 
\label{sqd-theory}
What happens if we replace the projective measurement used
to compute QD by a weak measurement?
Weak measurements, where the system-apparatus interaction
is weak, extract limited information from the system and
correspondingly disturb the system in a limited way.
Repeated weak measurements lead to a strong or a projective
measurement. 
The positive operator valued measure (POVM) 
corresponding to the weak measurement is defined through
the operators:
\begin{align}
P(x)&=&\sqrt{\frac{1-\tanh
x}{2}}\Pi_{\psi_{1}}^{\theta,\phi}+\sqrt{\frac{1+\tanh
x}{2}}\Pi_{\psi_{2}}^{\theta,\phi}\nonumber \\
\label{Weak_POVM} 
P(-x)&=&\sqrt{\frac{1+\tanh
x}{2}}\Pi_{\psi_{1}}^{\theta,\phi}+\sqrt{\frac{1-\tanh
x}{2}}\Pi_{\psi_{2}}^{\theta,\phi} 
\end{align}
where the strength of the weak measurement is parameterized
by the real parameter $x \geq 0$. The POVM operators satisfy
$P(x)^{\dagger}P(x)+P(-x)^{\dagger}P(-x)=I$.  For  $x=0$,
$P(0)$ reduces to $\frac{1}{\sqrt{2}}I$ \ie no  measurement
at all and in the case of $x\rightarrow\infty$, $P(x)$ and
$P(-x)$ reduce to the projectors
$\Pi_{\psi_{2}}^{\theta,\phi}$ and
$\Pi_{\psi_{1}}^{\theta,\phi}$, respectively (as defined in
Eq.~(\ref{orth_proj})).

After the  weak measurement, the state of the combined
system becomes:
\begin{equation}\label{state_after_weak_meas}
\begin{split}
\rho^{x}_{AB} &=\left(I\otimes P
\left(x\right)\right)\rho_{AB}\left(I\otimes P\left(x\right)\right)+ \\
&\quad \quad 
\left(I\otimes P\left(-x\right)\right)
\rho_{AB}\left(I\otimes P\left(-x\right)\right)
\end{split}
\end{equation}
The post-measurement state of
subsystem $A$ for each outcome can be written as
\begin{equation} 
\label{State_afterWM} 
\rho_{A \vert P(\pm x)}=
\dfrac{1}{p(\pm x)}\mathrm{Tr}_{B}\left[\left(I\otimes
P(\pm x)\right)\rho_{AB}\left(I\otimes
P(\pm x)\right)\right].
\end{equation}
Where $p(\pm x)$ are the probabilities for $P(\pm x)$.
The question now is: how can we define weak variants of quantum discord?
\subsubsection{Super quantum discord}
If we take Eq.~(\ref{QD2_1}) as the basic definition of
QD and replace the projective measurement with a weak POVM
(as defined above), we obtain 
a straightforward generalization of QD.
This can be written down by first computing 
the conditional entropy and classical information as:
\begin{align}
&S_{x}(\rho_{_{A\vert B}})=
p(x)S_{x} (\rho_{A} \vert P(x))
+p(-x)S_{x} (\rho_{A} \vert P(-x))\nonumber\\
&J_{x}(\rho_{A \vert B})=S(\rho_{_{A}})-S_{x}(\rho_{_{A\vert B}}).
\end{align}
In terms of the above, a weak variant of quantum discord 
can be defined which is called super quantum discord (SQD):
\begin{equation}
SQD(\rho_{AB})=I(\rho_{AB})-\max_{\lbrace \theta,\phi
\rbrace}J_{x}(\rho_{A\vert B})
\label{SQD}
\end{equation}
which depends on the measurement strength  $x$.  The
value of SQD is always greater than QD because $S_{x}(\rho_{A\vert
B})$ is always larger than $S(\rho_{A\vert B})$, as the weak
measurement is performed on subsystem $B$ while disturbing the
state weakly, and reveals less information about the subsystem
$A$.  In the limiting case of no measurement performed on
system $B$ at all, $S_{x}(\rho_{A\vert B}) \rightarrow
S(\rho_A)$ and SQD discord will be equal to the total
correlations.
\subsubsection{Weak Quantum Discord} 
\label{wqd-theory} 
SQD
has the feature that it is always larger than QD and has
hence
been found somewhat dissatisfying. A weak generalization of QD
is possible if we take Eq.~(\ref{QD2_2}) to be a 
fundamental definition of QD and replace the second term with
its weak equivalent~\cite{rulli}.  This process allows us to
define another weak variant of quantum discord
called weak quantum discord (WQD),
which for a two-qubit system can be written as:
\begin{equation}
WQD({\rho_{AB}})=I(\rho_{AB})-\max_{\{\theta,\phi \}}
I(\rho^{x}_{AB}) \end{equation} 
where $\rho^{x}_{AB}$ is the
density operator of the composite system after a
measurement of strength $x$ has been performed on the subsystem
$B$ as given in Eq.~(\ref{state_after_weak_meas}).
WQD as a quantifier has the nice property that it is always less than
QD and in the limit $x\rightarrow 0$ it approaches $0$, while for
$x\rightarrow \infty$ it approaches QD. In the strong measurement
limit, both SQD and WQD become the same and are equal to QD.  The
key observation which plays a role in these two
different generalizations of QD in the weak regime, is that
the two different equivalent expressions for QD do not remain
the same in the weak measurement regime. 
For interesting and more detailed interpretations of SQD and 
WQD, the reader is referred to Ref.~\cite{dieguez-qip-18}.
\section{Simulating Weak POVM Via a Phase Damping Channel} 
\label{phase_damping_and_weak_POVM}
A projective measurement collapses the state, thereby
killing all the off-diagonal terms (coherences) in the
density matrix, in the measurement basis~\cite{xiao-pla-06}.
The weak measurement formalism, as described by Aharonov,
Albert and Vaidmann (AAV)~\cite{aharonov-prl-88} utilizes
the weak interaction~\cite{neumann-book}. The weak interaction
couples the system weakly with the measuring device, and
therefore the state of the system retains its coherence
partially, even after the measurement. Later, Oreshkov and
Brun~\cite{oreshkov-prl-05} showed that any generalized
measurement can be modeled by a sequence of weak
POVMs and for the two-qubit case
are given in  Eq.(\ref{Weak_POVM}).

We consider two types of states to
investigate the behavior of 
the quantities SQD and WQD (defined in the
previous section) with respect to measurement
strength and to compare it with QD, namely, the Werner
states
and Bell-diagonal states. The two-qubit Werner states
are defined as:
\begin{equation}
\label{werner}
\rho_{_{AB}}^{ws}=z\vert\psi^{-}\rangle
\langle\psi^{-}\vert+\frac{1}{4}(1-z) I
\end{equation} 
where $1-z$ quantifies the amount of mixedness, $0 \leq z
\leq 1$ and
$
\vert \psi^{-}\rangle=
\frac{1}{\sqrt{2}}
(\vert 01\rangle -\vert 10\rangle)$.
The two-qubit Bell-diagonal
states~\cite{yinghua-optik-16}  are defined as:
\begin{equation}
\label{bell}
\rho_{_{AB}}^{bs}=\frac{1}{4}\left[I\otimes I 
+\sum_{i=1}^{3}c_{i}(\sigma_{i}\otimes\sigma_{i})\right]
\end{equation} 
where $(\sigma_{1},\sigma_{2},\sigma_{3})$ are the Pauli
matrices and $-1\leq c_{1},c_{2},c_{3}\leq 1$.

The evaluation of SQD and WQD in both the states involves an
optimization over all possible projectors by varying $\theta
\in [0,\pi]$ and $\phi \in [0,2\pi]$ as given in
Eq.(\ref{SQD}). The optimization gives the highest possible
classical correlations at $\theta=\pi$ and $\phi=\pi$ for
the Werner states as well for the
Bell-diagonal states.  On
substituting the optimal values of $\theta$ and $\phi$ into
Eq.(\ref{Weak_POVM}), the weak POVMs get simplified to:
\begin{eqnarray}
\!\!\!\!P(x)=\sqrt{\frac{1-\tanh{x}}{2}} \vert 0 \rangle \langle 0
\vert + \sqrt{\frac{1+\tanh{x}}{2}} \vert 1 \rangle \langle
1 \vert \nonumber \\
\!\!\!\!P(-x)=\sqrt{\frac{1+\tanh{x}}{2}}
\vert 0 \rangle \langle 0 \vert + 
\sqrt{\frac{1-\tanh{x}}{2}} \vert 1 \rangle \langle 1 \vert.
\nonumber \\
\label{qmi2}
\end{eqnarray}
A single-qubit mixed state $\rho$ on the Bloch sphere can be
expressed as:
\begin{equation}
\rho=\frac{1}{2}
\left(I+r_{x}\sigma_{1}+r_{y}\sigma_{2}+r_{z}\sigma_{3}\right)
\label{single_qubit}
\end{equation}
where $r_{x}$, $r_{y}$ and $r_{z}$ are the coordinates of
the Bloch vector and $I$ is the $2\times2$ identity matrix.
The effect of the simplified weak POVM given in 
Eq.(\ref{qmi2}) on the single-qubit state  $\rho$
can be readily computed and in the matrix form is written
as:
\begin{align}
\rho &=\dfrac{1}{2}
\begin{bmatrix}
1+r_{z}&r_{x}-ir_{y}\\
r_{x}+ir_{y}&1-r_{z}\\
\end{bmatrix}\nonumber \\
&\;\;\;\;\;\;\;\;\;\;\;\;\;\;\;\;\;\;\;\;\;\;\;\;
\Downarrow\nonumber\\
\rho_{_{wm}}' &=\dfrac{1}{2}
\begin{bmatrix}
1+r_{z}&(r_{x}-ir_{y})\sech x\\
(r_{x}+ir_{y})\sech x&1-r_{z}.\\
\end{bmatrix}
\end{align}
It is clear from the post weak-measurement state
$\rho^{\prime}_{wm}$ that the off-diagonal terms are a
monotonically decreasing function of the measurement
strength $x$, leading to decoherence. The extent to which the
weak measurement decoheres the state $\rho$ depends on the
measurement strength $x$.

We now turn to the PD channel, which  causes loss of
coherence and leads to the decay of the off-diagonal terms
of the density matrix and can be described by a completely
positive trace preserving map described through the Kraus
operators~\cite{ruskai-laa-02}:
\begin{eqnarray}
&\rho^{\prime}_{PD}=E_{0} \rho E_{0}^{\dagger}+E_{1} \rho E_{1}^{\dagger}
\nonumber \\
&
E_{0}=\frac{1+\sqrt{1-\lambda}}{2}I+\frac{1-\sqrt{1-\lambda}}{2}\sigma_{3}\nonumber
\\
&E_{1}=\frac{\sqrt{\lambda}}{2}I-\frac{\sqrt{\lambda}}{2}\sigma_{3}
\label{kraus_rep}
\end{eqnarray}
where the parameter $\lambda \in [0,1]$ represents the
strength of the PD channel. 
\begin{figure}[t]
\includegraphics[angle=0,scale=1]{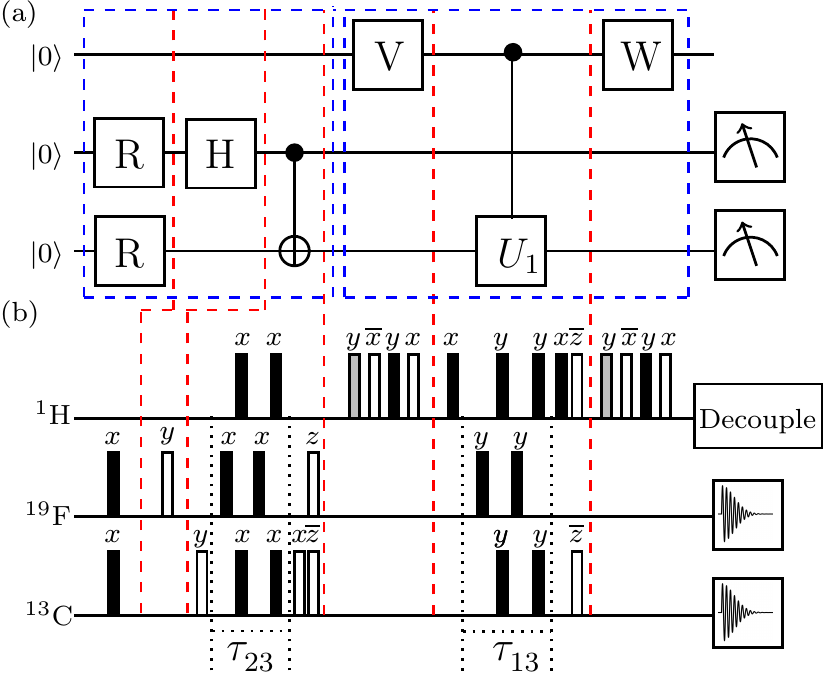}
\caption{(a) 
The quantum circuit in the left block creates a Bell-diagonal 
state and a Werner state ($\rho_{0}\otimes$
$\rho_{\psi}$) 
on two qubits ( $ \rho_{\psi} $), with the
third qubit ($ \rho_{0} $) acting as ancillary; H denotes a
Hadamard gate, R is a NOT gate in the case of the Werner
state and $I_2$ identity operation (no operation) in the
case of the Bell-diagonal 
state.  The quantum circuit in the
right block implements a phase damping channel on one of the
qubits of hence prepared two-qubit state using an ancillary
qubit. (b) NMR pulse sequences corresponding to the quantum
circuits where the unfilled rectangles denote
$\frac{\pi}{2}$ radiofrequency (rf) pulses, the filled rectangles denote
$\pi$ rf pulses and the shaded rectangles denote $\theta$ rf
pulses where $\theta=-2
\sin^{-1}\sqrt{\frac{1-\sqrt{1-\lambda}}{2}}$, and $\lambda$
is the strength of the PD channel, lying between $0$ and
$1$.  The phase of the rf pulse is given above each pulse
and a bar over a phase represents negative phase. The free
evolution time intervals $\tau_{12}$ and $\tau_{23}$ are
given by $1/(2 \textrm{J}_{12})$ and $1/(2 \textrm{J}_{23})$
respectively, where $\rm{J}_{ij}$ represents the scalar
coupling strength between qubits $i$ and $j$.} 
\label{ckt_wd} 
\end{figure}
\begin{figure}[t]
\includegraphics[angle=0,scale=1]{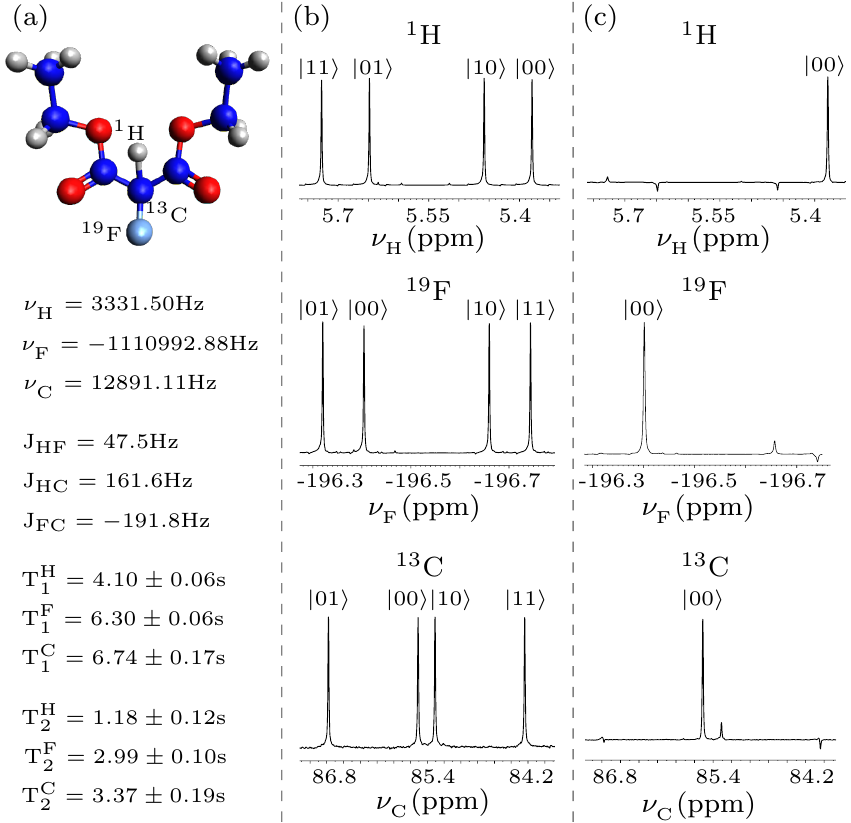}
\caption{$\left(a\right)$ Molecular structure of
$^{13}$C-labeled diethylfluoromalonate with the three qubits
labeled as $^{1}$H, $^{19}$F and $^{13}$C. NMR parameters
i.e. the chemical shift $\nu_{i}$ (in Hz) of each nuclear
spin,
spin-spin
coupling between them $\textrm{J}_{ij}$ (in Hz), spin-lattice
relaxation times T$_{1}^{i}$ and spin-spin relaxation times
T$_{2}^{i}$  (in seconds). NMR spectrum of (b) thermal
equilibrium state obtained after a $\frac{\pi}{2}$ readout
pulse and (c) pseudopure state. The resonance lines of each
qubit in the spectra are labeled by the corresponding
logical states of the other qubits.}
\label{mol} 
\end{figure}
The action of the PD channel on a general one-qubit state in
the matrix form is given by:
\begin{align*}
\rho &=\frac{1}{2}
\begin{bmatrix}
1+r_{z}&r_{x}-ir_{y}\\
r_{x}+ir_{y}&1-r_{z}\\
\end{bmatrix}\\
&\;\;\;\;\;\;\;\;\;\;\;\;\;\;\;\;\;\;\;\;\;\;\;\; \Downarrow\\
\rho_{_{PD}}' &=\frac{1}{2}
\begin{bmatrix}
1+r_{z}&(r_{x}-ir_{y})\sqrt{1-\lambda}\\
(r_{x}+ir_{y})\sqrt{1-\lambda}&1-r_{z}.\\
\end{bmatrix}
\end{align*}
The effect of the PD channel is similar to the weak POVM on
a single-qubit state, wherein the off-diagonal terms are
diminished.  Since both `$\sech x$' and `$\sqrt{1-\lambda}$'
are monotonically decreasing functions, they can be mapped
onto each other with an appropriate scaling factor. 
Therefore, the action of the PD channel is in one-to-one
correspondence with weak POVM described in
Eq.~(\ref{Weak_POVM}).
\begin{figure}[t]
\includegraphics[angle=0,scale=1]{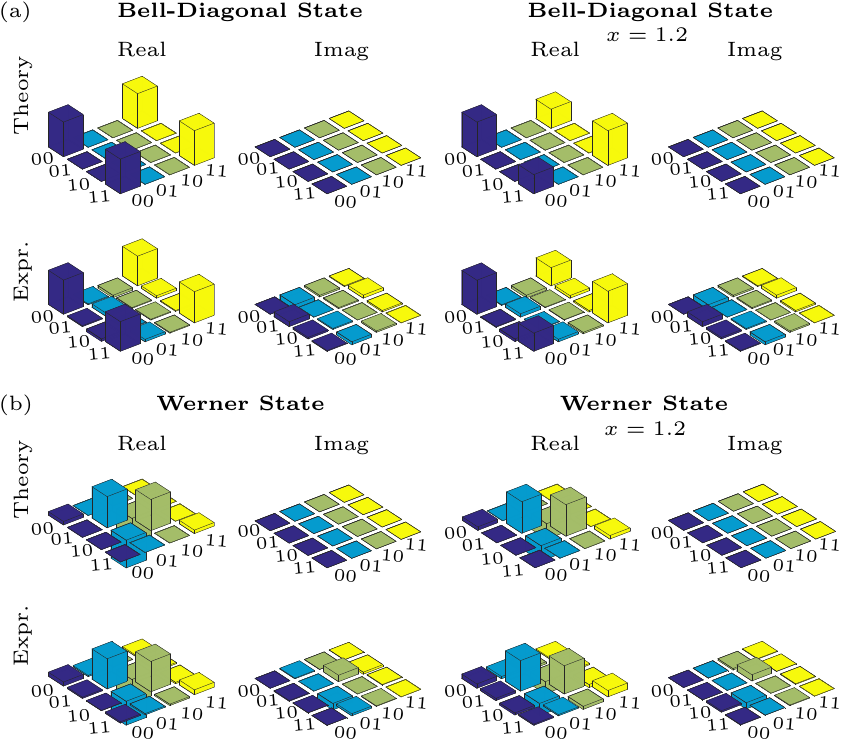}
\caption{Real and imaginary parts of theoretically expected
and the experimentally reconstructed tomographs of (a)
Bell-diagonal state 
and (b) Werner state before (left) and after (right) PD
channel implementation at a measurement strength $x=1.2$. }
\label{tomo} 
\end{figure}

In order to implement the PD channel we follow an indirect
approach~\cite{xin-pra-17}, wherein non-unitary operators
can be thought of as unitary operations on an extended
quantum system built upon the Duality Quantum Computing
(DQC) framework~\cite{long-ctp-06}. This framework requires
an ancillary qubit. It has been
demonstrated~\cite{xin-pra-17} that the Kraus operator
${E_{k}}$ describing the non-unitary transformation
corresponding to the PD channel can be efficiently
implemented on a qubit if the unitary operators $V$, $W$, $U_{0}$ and
$U_{1}$ can be found in the two-qubit space, where $V$ and
$W$ act on the ancilla qubit, and $U_0$ and $U_1$ act on the
target qubit controlled by the
ancilla qubit. The operators need to satisfy:
\begin{equation}
E_{k}(k=0,1)=\sum_{i=0}^{1}W_{ki}V_{i0}U_{i}
\end{equation}
where $W_{ki}$ is $ki$th element of $W$ operator, $V_{i0}$ is
an element of 
the first column of $V$ operator and $U_{i}$ is the
controlled operator. A comparison with the
decomposition of Kraus operators of the PD channel is given in 
Eq.(\ref{kraus_rep}).
The unitary operators $V$, $W$, $U_{0}$ and
$U_{1}$ can be evaluated as:
\begin{eqnarray}
\label{unitary_op}
&&U_{0}=I,\quad U_{1}=\sigma_{3},\nonumber \\
&&V=W=\frac{1}{2}
\begin{bmatrix}
\sqrt{\frac{1+\sqrt{1-\lambda}}{2}}&\sqrt{\frac{1-\sqrt{1-\lambda}}{2}}\\
\sqrt{\frac{1-\sqrt{1-\lambda}}{2}}&-\sqrt{\frac{1+\sqrt{1-\lambda}}{2}}
\\
\end{bmatrix}.
\end{eqnarray}

The quantum circuit to implement the PD channel is shown in
Fig.~\ref{ckt_wd}(a), 
where Werner states and Bell-diagonal states
(Eq.(\ref{werner}-\ref{bell}))
states are created on
qubits 2 and 3, qubit 1 acts as an ancilla, and the  
the PD channel acts on qubit 3.
The strength of the PD channel is controlled by the $V$ gate. 
The effect of the PD channel can be 
evaluated by tracing out
the ancillary qubit. Physically this was achieved by
performing the measurements on the state of the two-qubit
subsystem consisting of qubits 2 and 3, while ignoring the
qubit 1, as is shown in Fig.~\ref{ckt_wd}.  

We are now ready to experimentally investigate the
behavior of SQD and WQD 
by varying the
measurement strength. It is important to mention here that
in an NMR set up the measurement is already weak (termed as
an ensemble weak measurement) since the interaction of the
measuring rf coil with the nuclear spins is
weak~\cite{lee-apl-06}. However, we are not using that weak
measurement here. Our weak measurement is simulated in a
controlled way by the PD channel, which is implemented with the
help of the ancilla qubit.
\section{Experimental Implementation of a Weak POVM} 
\label{Experiment} 
As discussed earlier in
Sec.~\ref{phase_damping_and_weak_POVM}, weak measurements
can be mapped onto a PD channel and the strength of the weak
measurement can be varied by tuning the strength of the PD
channel. 
For the experimental realization on an NMR quantum
processor, we realize the three qubits 
as the three spin-1/2 nuclei of
$^{13}$C-labeled diethyl fluoromalonate dissolved in
acetone-D6. The $^{1}$H, $^{19}$F and $^{13}$C nuclear spins
are labeled as the first, second and third qubit,
respectively. It should be noted here that two-qubit system
was simulated by $^{19}$F and $^{13}$C nuclear spins while
$^{1}$H spin was utilized as the ancillary qubit. The molecular
structure along with relevant experimental parameters and
corresponding NMR spectrum of the thermal equilibrium state
are shown in Figs.~\ref{mol}(a) and (b) respectively.
The Hamiltonian for a three-qubit system in a rotating frame
under the weak approximation~\citep{ernst-book-90} is given
by: 
\begin{equation}
\label{Hamiltonian}
H=-\sum_{i=1}^{3}\nu_{i}I_{z}^{i}+\sum_{i>j,i=1}^{3}
\textrm{J}_{ij}I_{z}^{i}I_{z}^{j}
\end{equation} 
where $i,j=$1,2 and 3 labels the qubit,
$v_{i}$ represents the chemical shift of the respective
nuclei, $\textrm{J}_{ij}$ is the scalar coupling constant between the
$i^{th}$ and $j^{th}$ nuclear spins and $I_{z}^{i}$ denotes
the $z$ component of the spin angular momentum operators for
the $i^{th}$ nucleus. We used the 
spatial averaging technique~\cite{cory-physD-98,mitra-jmr-07} 
to achieve the initial 
three-qubit pseudopure state (PPS) $\vert000\rangle$ from
the thermal equilibrium state, with the density operator $\rho_{000}$ 
given by: 
\begin{equation}
\label{PPS}
\rho_{000}=\frac{(1-\epsilon)}{8}I +\epsilon \vert 000
\rangle \langle 000 \vert 
\end{equation} 
where $\epsilon\sim 10^{-5}$ represents the spin polarization at room
temperature and $I$ is the 8$\times$8 identity operator.
The identity operator does not evolve and the measurable 
NMR signal can be attributed to the deviation
density matrix. The NMR spectrum of the three-qubit PPS is shown
in Fig.~\ref{mol}(c). The experimentally prepared PPS was
tomographed using full quantum state 
tomography~\citep{leskowitz-pra-04}. The state fidelity was found to
be $0.981 \pm 0.006$ and was computed using the Uhlmann-Jozsa
measure~\citep{uhlmann-rpmp-76,jozsa-jmo-94}:
\begin{equation}\label{fidelity}
\rm F=\left[Tr\left( \sqrt{\sqrt{\rho_{{\rm th}}}\rho_{{\rm
ex}} \sqrt{\rho_{{\rm th}}}}\right)\right]^2
\end{equation}
where $\rho_{\rm th}$ and $\rho_{\rm ex}$ denote
the theoretical and experimental density
operators, respectively and F is normalized 
using $ \rm F \rightarrow 1 $ as $\rho_{\rm ex} \rightarrow
\rho_{\rm th}$.
All the experimental density matrices were reconstructed by
performing full quantum state tomography
\citep{leskowitz-pra-04} using a set of seven preparatory
pulses $\left\lbrace III, XXX, IIY, XYX, YII, XXY, IYY
\right\rbrace$, where $I$ represents `no-operation' and $
X(Y) $ denotes local $ \frac{\pi}{2} $ unitary rotation with
phase $ x(y) $ which is implemented by applying
a spin-selective $\frac{\pi}{2}$ pulse. We performed all the
experiments at room temperature 293K on a Bruker Avance-III
600 MHz FT-NMR spectrometer equipped with a QXI probe.

The quantum circuit to implement a weak measurement,
simulated by a PD channel, is shown in Fig.~\ref{ckt_wd}(a).
The left block in the circuit creates a Werner or a
Bell-diagonal
state $(\rho_{0}\otimes
\rho_{\psi})$ on two
qubits ($ \rho_{\psi} $) with the third qubit ($  \rho_{0} $) acting as ancillary;  R is a NOT
gate in the case of the Werner state and a `no-operation'
for a Bell-diagonal 
state.  The right block in the circuit
depicts a PD channel acting on the one of the qubits of two-qubit state using
an ancillary qubit, and the strength of the PD channel
being controlled by a local rotation achieved by the $V$
gate acting on the ancillary qubit.

\begin{table}[h]
\caption{\label{r_table}
Experimental (Exp) results of 
weak measurement strength $x$ varied in 
a Werner (W) and a Bell-diagonal (BD) 
state while implementing the PD channel.}
\begin{center}
\begin{tabular}{ccc}
\hline
\hline
Theory $x$ & Exp $x$ (W) & Exp $x$ (BD) \\
\hline
\hline
0.00 & 0.091 $\pm$ 0.047 & 0.069 $\pm$ 0.036 \\
0.34 & 0.503 $\pm$ 0.018 & 0.373 $\pm$ 0.002 \\
0.55 & 0.666 $\pm$ 0.015 & 0.548 $\pm$ 0.005 \\
0.75 & 0.831 $\pm$ 0.016 & 0.721 $\pm$ 0.005 \\
0.95 & 1.016 $\pm$ 0.016 & 0.907 $\pm$ 0.004 \\
1.20 & 1.215 $\pm$ 0.020 & 1.103 $\pm$ 0.007 \\
1.50 & 1.479 $\pm$ 0.024 & 1.350 $\pm$ 0.007 \\
1.75 & 1.819 $\pm$ 0.045 & 1.667 $\pm$ 0.018 \\
2.00 & 2.122 $\pm$ 0.046 & 1.937 $\pm$ 0.018 \\
2.50 & 2.454 $\pm$ 0.062 & 2.213 $\pm$ 0.026 \\
3.00 & 3.242 $\pm$ 0.151 & 2.795 $\pm$ 0.042 \\
3.50 & 3.758 $\pm$ 0.095 & 3.448 $\pm$ 0.069 \\
4.00 & 4.107 $\pm$ 0.167 & 4.259 $\pm$ 0.123 \\
4.50 & 4.402 $\pm$ 0.062 & 4.575 $\pm$ 0.069 \\
5.00 & 4.839 $\pm$ 0.137 & 5.030 $\pm$ 0.180 \\
\hline
\hline
\end{tabular}
\end{center}
\end{table}

Experimentally, we prepared a Werner state with mixedness
strength $z=0.8$.  The second term on the RHS of
Eq.(\ref{werner}) is a singlet state and was created
experimentally, followed by quantum state
tomography (QST). To obtain the Werner state
of a desired $z$ value the QST generated singlet state was
numerically added to the identity and thus, a Werner state
with fidelity $0.990 \pm 0.001$ was created.  Next the
Bell-diagonal 
state with $c_{1}=1$, $c_{2}=-1$ and $c_{3}=1$ was
prepared experimentally with fidelity $0.980 \pm 0.002$.
The next step is to perform a weak measurement on
the second qubit (treated as a subsystem of the two-qubit
system).  As described in
Sec.~\ref{phase_damping_and_weak_POVM}, the weak measurement
is simulated by the PD channel and the strength of PD
channel is controlled by the real parameter $\lambda$ which
is related to weak measurement strength $x$ as
$\lambda=1-\sech^{2}{x}$. We implemented the PD channel on the
one of the qubits of the prepared Werner state using the 
ancillary qubit and the strength of the PD channel was
increased corresponding to weak measurement strength $x$ as
shown in Table-\ref{r_table}.  
\begin{figure}[t]
\includegraphics[angle=0,scale=1]{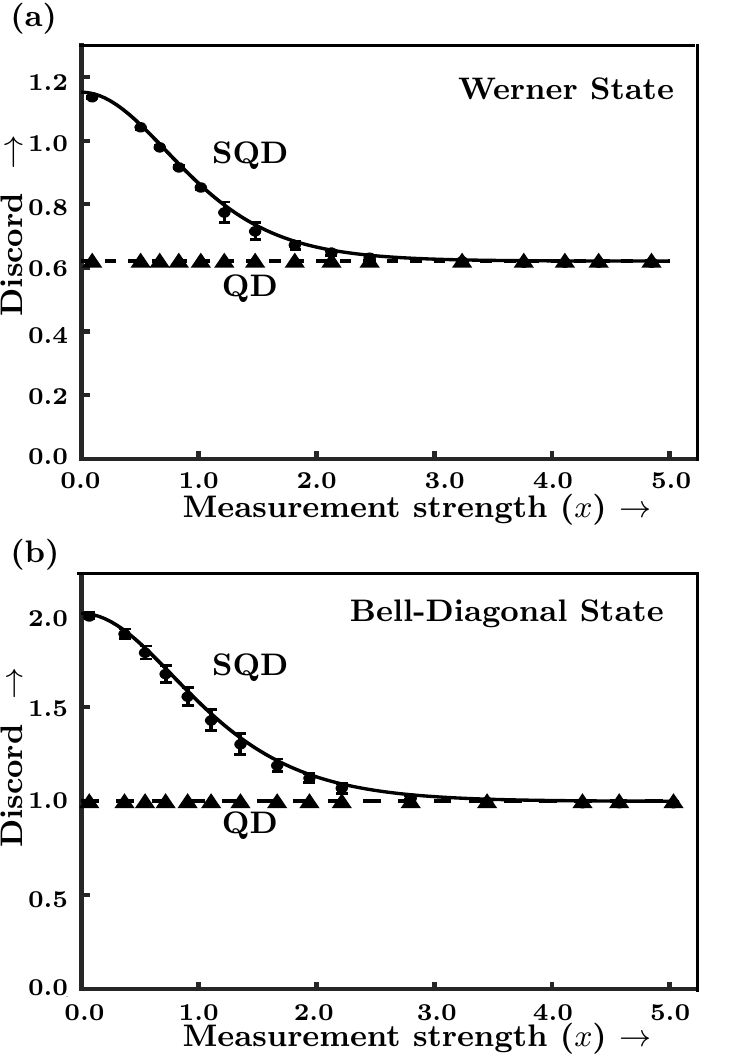}
\caption{Plot
of SQD and QD with measurement
strength $(x)$ for the two-qubit experimentally prepared (a)
Werner state and (b) Bell-diagonal state.
The solid and dashed lines represent the theoretically
computed values while the filled circles and triangles represent the
experimental values of SQD and QD, respectively.}
\label{sqd-figure}
\end{figure}
\begin{figure}[t]
\includegraphics[angle=0,scale=1]{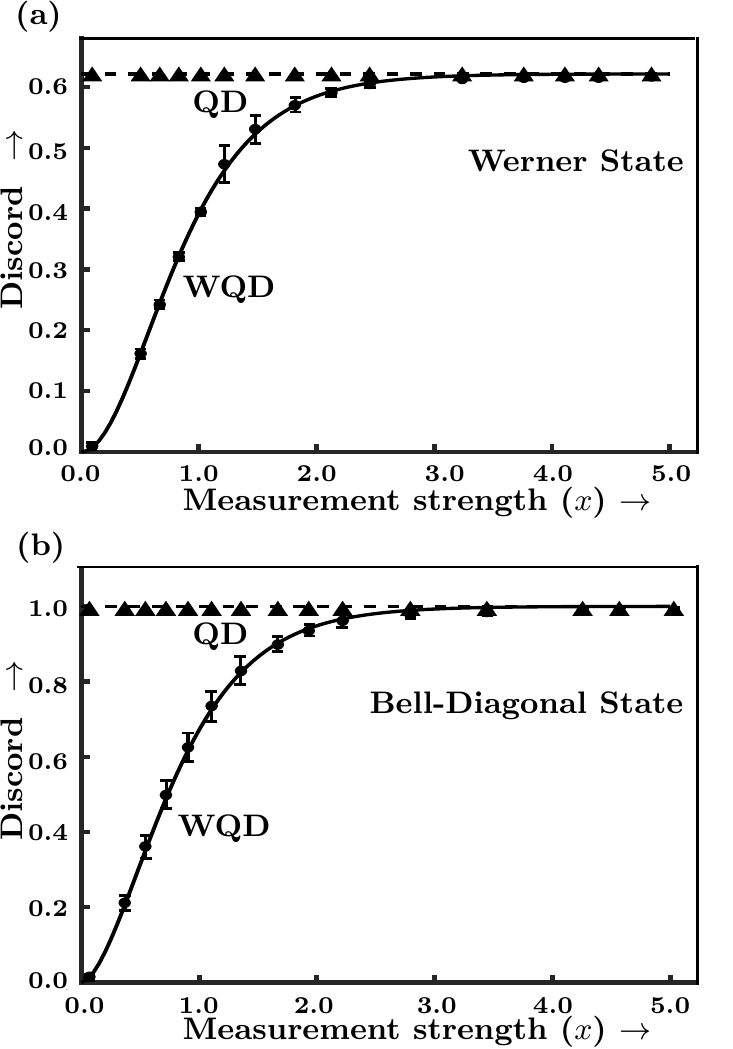}
\caption{Plot
of WQD and QD with measurement
strength $(x)$ for the two-qubit experimentally prepared (a)
Werner state and (b) Bell-diagonal state.
The solid and dashed lines represent the theoretically
computed values while the filled circles and triangles represent the
experimental values of WQD and QD, respectively.}
\label{wqd-figure}
\end{figure}
We performed a similar experiment for the
Bell-diagonal states, which
was directly prepared from the left block in the circuit as
depicted in Fig.~\ref{ckt_wd}(a) and the PD channel was
implemented with increasing weak measurement strength $x$ as
shown in Table-\ref{r_table}. All the experimentally
prepared three-qubit states were tomographed before and
after implementing PD channel. Both the two-qubit Werner and
Bell-diagonal 
states were reconstructed utilizing QST and tracing out
the ancillary qubit. The tomograph of one such experimentally
reconstructed density matrix of both initially prepared
states is shown on the LHS of Fig.~\ref{tomo}(a) and
Fig.~\ref{tomo}(b), respectively. The tomographs on the RHS
of Fig.~\ref{tomo}(a) and (b) depicts the states after
the action of the PD channel corresponding to the weak
measurement strength $x=1.2$.

We investigated theoretically and experimentally, the
behavior of SQD, WQD, and QD present in both Werner and
Bell-diagonal states while increasing the measurement
strength $x$, and the results are plotted in
Fig.~\ref{sqd-figure} and Fig.~\ref{wqd-figure},
respectively. Our results show that in both types of states,
the value of SQD is always greater than QD and 
is maximum at zero measurement strength, which implies
that the measured state is undisturbed.  
On the other hand, as a quantifier WQD is never greater than QD and 
allows the researcher to navigate the region between $x
\longrightarrow 0$  when no measurement is performed (and
quantum correlations are intact)  and
$x \longrightarrow \infty$ when a projective measurement is 
performed (and quantum correlations are destroyed).
Furthermore, as evidenced from Fig.~\ref{sqd-figure}
and Fig.~\ref{wqd-figure}, both SQD and WQD approach QD
as the measurement strength increases. Our experimental
results are hence in consonance with the theoretically expected
behavior of the quantifiers of discord for the case of
a weak measurement.
\section{Concluding Remarks} 
\label{conclusion}
We have implemented a weak POVM, exploiting its relationship
with the PD channel, on an NMR quantum information
processor.  The noise induced by the PD channel has been
exploited to mimic the disturbance introduced by a weak
measurement process.  The weak POVM was experimentally
applied to find the SQD and WQD in two classes of bipartite quantum
states, namely, the Bell-diagonal state and the Werner state. The
SQD and WQD were contrasted against QD 
and it was observed that both these quantities
converge to QD as the strength of the measurement increases.
The monotonicity of SQD and WQD was also confirmed. 
Our results could be useful for iterative experimental
information processing protocols which seek to disturb the
state only slightly.  Although the interpretation of 
weak variants of QD
remains elusive, this work opens up the possibility of
further experimental investigations on 
such variants, which could
potentially exploit correlations beyond
QD~\cite{dakic-np-12,cavalcanti-pra-11}.

\vspace*{1cm}
\noindent{\bf Declaration of Interest:} The authors declare
that they have no conflict of interest.\\

\vspace*{1cm}
\noindent{\bf Acknowledgments}
All the experiments were performed on a Bruker Avance-III
600 MHz FT-NMR spectrometer at the NMR Research Facility of
IISER Mohali.

\end{document}